# A DISTRIBUTED TRUST MANAGEMENT FRAMEWORK FOR DETECTING MALICIOUS PACKET DROPPING NODES IN A MOBILE AD HOC NETWORK

Jaydip Sen

Innovation Lab, Tata Consultancy Services Ltd.,
Bengal Intelligent Park, Salt Lake Electronic Complex, Kolkata India
Jaydip.Sen@tcs.com

## ABSTRACT

*In a multi-hop mobile ad hoc network (MANET) mobile nodes communicate with each other forming a cooperative radio network. Security remains a major challenge for these networks due to their features of open medium, dynamically changing topologies, reliance on cooperative algorithms, absence of centralized monitoring points, and lack of any clear lines of defense. Most of the currently existing security algorithms designed for these networks are insecure, in efficient, and have low detection accuracy for nodes' misbehaviour. In this paper, a new approach has been proposed to bring out the complementary relationship between key distribution and misbehaviour detection for developing an integrated security solution for MANETs. The redundancy of routing information in ad hoc networks is utilized to develop a highly reliable protocol that works even in presence of transient network partitioning and Byzantine failure of nodes. The proposed mechanism is fully co-operative, and thus it is more robust as the vulnerabilities of the election algorithms used for choosing the subset of nodes for cooperation are absent. Simulation results show the effectiveness of the proposed protocol.*

## 1. INTRODUCTION

Although the security objectives of both ad hoc networks and traditional networks are considered to be the same such as availability, confidentiality, integrity, authentication, and non-repudiation, the security issues involved in ad hoc networks are quite different due to the 'mobile' and 'ad hoc' constraints, i.e. limited resources for computation and communication, dynamic network topology as well as the mobility of the hosts. In traditional networks, most trust evidences are generated via potentially lengthy assurance processes, distributed off-line, and assumed to be valid on a long term. In contrary, few of these characteristics of trust relations and trust evidences are prevalent in MANETs. Cryptographic primitives such as authentication and key distribution are the usual mechanisms used for implementing security in MANETs. However, these schemes cannot provide security against some attacks such as packet dropping attack by malicious nodes in the network.

In general, uncooperative nodes in MANETs may be of two types: *malicious nodes* and *selfish nodes*. The nodes belonging to the first category are either faulty and therefore cannot follow a protocol, or are intentionally malicious and try to attack the system. The problems created by these nodes need to be addressed at multiple layers, for example, using spread-spectrum encoding to avoid interference over the communication channel, using a reputation system to identify the malicious system, and subsequently avoiding or penalizing such nodes. A selfish node, on the other hand, is an economically rational node whose objective is to maximize its own welfare, which is defined as the benefit of its actions minus the cost of its actions. Since forwarding a message will incur a cost, a selfish node will need incentive for doing it.

There are several approaches towards designing security mechanisms for MANETs [1][2][3][4]. The significant efforts done so far are mainly in the adaptation from the existing distributed trust model to the ad hoc trust model. One approach of establishing trust among the nodes in a MANETs is by detecting misbehaving nodes that maliciously drop packets. These malicious nodes can be detected by utilizing the concept of reputation [5][6]. The reputation of a node refers to the perception that another node has about its intention and activities. Reputation is a tool for motivating cooperation among nodes so as to ensure that most of them exhibit good behaviour in their activities. Each node in a network is assigned a reputation value as computed jointly by its neighbors. The higher the reputation value of a node the more trustworthy the node is. However, there are several issues with the reputation systems. First, there are no formal specifications and analyses for the types of incentives provided by such systems. Second, these systems do not consider the possibility that even selfish nodes can collude with each other in order to maximize their welfare. Third, some of these schemes depend on broadcast nature of wireless networks for monitoring. Such monitoring, however, may not always be possible due to asymmetric links when the nodes use power control. Further, directional antennas, which have gained momentum in the filed of wireless networks, will also make monitoring hard.

Another approach to provide incentive is to use credit or virtual currency. Buttyan and Hubaux proposed an elegant solution of this type in [7], and then presented an improved set of results based on credit counters [8]. In both these proposals, a node receives one unit of credit for forwarding a message of another node, and such credits are deducted from the sender (or the destination). However, both these proposals require a tamper-proof hardware at each node to ensure correct credit and debit of *nuglets*.

In this paper, a reputation- and trust-based security framework for MANETs has been proposed that detects packet-dropping attacks launched by malicious and selfish nodes. The mechanism depends on a trust model that is based on the reputation of different nodes. MANETs are also prone to the following security threat: a node could be tempted to discard its initial identity and re-enter the network in disguise in environments where users are punished for their selfish or malicious behavior. A solution to this problem is also suggested by establishing a *univocal relation* between a physical device and the identity it claims as it first enters the network.

The rest of the paper is organized as follows: Section 2 discusses some of the major security vulnerabilities in MANETs. Section 3 presents some of the existing trust- and reputation-based schemes designed for MANETs. Section 4 describes the trust setup of the network in which the proposed protocol would work. Section 5 presents the proposed security scheme. Section 6 makes a security analysis of the scheme. Section 7 presents the simulation results, and Section 8 concludes the paper.

## 2. SECURITY VULNERABILITIES IN MANETS

Due to absence of any infrastructure, MANETs have a number of security vulnerabilities. The use of wireless links renders the network highly susceptible to link attacks including passive eavesdropping, active interfering, leakage of secret information, data tampering, impersonation, message replay, message distortion, and denial-of-service. Secondly, the nodes have relatively poor physical protection and are susceptible to being captured, compromised, and hijacked. Thirdly, the absence of infrastructure and authentication facilities in MANETs impedes the usual practice of establishing a line of defense, distinguishing the nodes as trusted and non-trusted categories. Furthermore, the nodes in a MANET are stingy in communication due to slower links, limited bandwidth, higher cost, and battery power constraints and therefore, computationally intensive security algorithms cannot be used for securing such a network. An additional problem in MANETs is the security    vulnerability of the routing protocols. A set of nodes in a MANET may be compromised in such a way that it may not be possible to detect their malicious behavior easily. Such nodes may generate new routing messages to advertise non-existent links, provide incorrect link-state information, and flood other nodes with routing traffic, thus inflicting Byzantine failure in the network.

These inherent vulnerabilities of MANETs cannot be prevented by intrusion prevention techniques such as encryption and authentication as these mechanisms are not capable of defending attacks launched by compromised nodes. A compromised node possesses the secret key and hence can encrypt and decrypt messages just like a legitimate node. Therefore, to build a secure MANET, deployment of an intrusion detection and response mechanism is needed so that any attack is detected at the earliest possible time and appropriate action taken to mitigate the attack.

## 3. RELATED WORK

There have been different approaches to define trust. Trust, in general, is a directional relationship between two entities and plays a major role in building a relationship between nodes in a network. Even though trust has been formalized as a computational model, it still means different things for different research communities. For example, the problem of defining trust metrics and trust relationship has been extensively studied for public key authentication [9][10][11], electronic commerce [12], as well as in P2P networks [13]. In some of these schemes, discrete or continuous numerical values are assigned to measure the level of trust [10][11][13]. For example, in [10], an entity's opinion about the trustworthiness of a certificate is described by a continuous value in [0,1]. In [11], a triplet in $[0, 1]^3$ is assigned to measure the trustworthiness where the elements in the triplet represent belief, disbelief, and uncertainty respectively. In [10], discrete integers are used. In [14] failed and selfish behaviors in ad hoc networks are studied.

The *reputation* of an entity has been defined as an expectation of its behavior based on other entities' observations or information about the entity's past behavior within a specific context at a given time [15]. In case of a MANET, the reputation of a node refers to how good the node is in terms of its contribution to routing activities in the network.

The *distributed trust model* proposed by Abdul-Rahman et al. uses a recommendation protocol to exchange trust-related information [16]. The trust relationships are assumed to be unidirectional between two entities. The recommendation protocol works by requesting a trust value in a target node with respect to a particular classifier. When the response arrives, an evaluation function is used to compute the overall trust value in the target. The protocol also allows recommendation refreshing and revocation. The model is suited for systems that are less formal and temporary in nature, e.g., some ad hoc commercial transactions.

The *resurrecting duckling* security protocol proposed by Stajano et al. is particularly suited for devices without display and embedded devices that are too weak for public-key operations [17]. The authentication problem is solved by a secure and transient association between two devices establishing a master-slave relationship. The association is secure because the master and the slave share a common secret, and it is transient because it can be terminated by the master at any point of time.

Kong et al. have proposed a trust building scheme for ad hoc networks that is similar to the *pretty good privacy* (PGP) web of trust concept [2]. However, unlike PGP it has no central certificate directory. In order to find the public key of a remote user, a local user makes use of the *Hunter algorithm* [3] on the merged certificate repository to build certificate chain(s).

Eshenauer et al. have proposed a trust establishment mechanism for MANETs [18]. In this scheme, a node in the network can generate trust evidence about any other node. When a *principal* generates a piece of trust evidence, it signs the evidence with its own private key, specifying the lifetime and makes it available to other through the network. A principal node may revoke a piece of evidence it produced by generating a revocation certificate for that piece of evidence and making it available to others, at any time before the evidence expires. A principal can get disconnected after distributing trust evidence. Similarly, a producer of trust evidence does not have to be reachable at the time its evidence is being evaluated. Evidences can be replicated across various nodes to guarantee availability. Although the scheme seems conceptually sound, the authors have provided no details about any performance evaluations.

Among the more recent works, Repantis et al. have proposed a decentralized trust management middleware for ad hoc, peer-to-peer networks based on reputation of the nodes [19]. In this scheme, the reputation information of each peer is stored in its neighborhood and piggybacked on its replies.

In the trust-based data management scheme proposed by Patwardhan et al., mobile nodes access distributed information, storage and sensory resources available in pervasive computing environment [20]. The authors have taken a holistic approach that considers data, trust, security, and privacy issues and utilizes a collaborative mechanism that provides trustworthy data management platform in a MANET.

Sun et al have presented a framework to quantitatively measure trust, model trust propagation, and defend trust evaluation system against malicious attacks [21]. The attacks against trust evaluation are identified and defense techniques have been proposed.

Baras and Jiang have presented a trust management scheme for self-organized ad hoc networks, where the nodes share trust information only with their neighbors [22]. For establishing and maintaining trust among the neighbors, the authors have proposed a voting mechanism.

Chang et al. have proposed a trust-based scheme for multicast communication in a MANET [23]. In a multicast MANET, a sender node sends packets to several receiving nodes in a multicast session. Since the membership in a multicast group changes frequently in a MANET, the issues of supporting secure authentication and authorization in a multicast MANET is very critical. The proposed scheme involves a two-step secure authentication method. First, an ergodic continuous Markov chain is used to determine the trust value of each one-hop neighbor. Second, a node with the highest trust value is selected as the *certificate authority* (CA) server. For the sake of reliability, the node with the second highest trust value is selected as the backup CA server. The analytical trust value of each mobile node is found to be very close to that observed in the simulation under various scenarios. The speed of the convergence of the analytical trust value shows that the analytical results are independent of the initial values and the trust classes.

Sun et al. have presented trust as a measure of uncertainty [24]. Using the theory of entropy, the authors have developed a few techniques to compute trust values from certain observations. In addition, trust models – entropy-based and probability-based, presented to solve the concatenation and multi-path trust propagation problems in a MANET.

Sen et al. have proposed a self-organized trust establishment scheme for nodes in a large-scale MANET in which a trust initiator is introduced during the network bootstrapping phase [25]. It has been proven theoretically and shown by simulation that the new nodes joining the network have high probability of successful authentication even when a large proportion of the existing nodes leave the network at any instant of time. A distributed intrusion detection system has been proposed in [26], where local anomaly detection is utilized to make a more accurate network-wide (i.e. global) detection using a cooperation detection algorithm among the nodes.

*Cooperation Of Nodes-Fairness In Dynamic Ad-hoc NeTworks* (CONFIDANT) is a security model based on selective altruism and utilitarianism proposed by Buchegger and Boudec to make misbehaviour unattractive in MANETs [27]. It is a distributed, symmetric reputation model that uses both first-hand and second-hand information for computation of reputation values. CONFIDANT uses *dynamic source routing* (DSR) protocol for routing and assumes that promiscuous mode of operation is possible. The misbehaving nodes are punished by isolating them from accessing the network resources.

The proposed protocol in this paper has many similarities with the CONFIDANT protocol. However, the metrics for computing the reputation of a node in the proposed protocol are different from those used in CONFIDANT. The proposed protocol takes into account the historical data of the reputation of the nodes which makes the computed reputation values more robust. In contrast to the approach followed in CONFIDANT, the proposed mechanism broadcasts the reputation information to all neighbors of a node thereby making the protocol more reliable and fault-tolerant and hence more secure.

## 4. TRUST SETUP

In an infrastructureless network such as a MANET, before a node is deployed, it has to generate a pair of public/private keys. For proper authentication of a node, there must be an association between the identity of the node and its public key. Public key certificates are usually used for this purpose. For use of certificates, an identifier of the node (e.g., IP address, MAC address etc.) must be defined which will be associated with its certificate. A malicious node may be tempted to discard its initial identity when its rating falls below a certain threshold. In reputation-based systems, when a node is isolated from the network because of its low reputation value, one way for the node to re-enter the network is to start from beginning in disguise. Therefore, a security scheme for MANETs should not only take care of bootstrapping authentication of each node, but it should also avoid delivery of two or more certificates that link different identifiers to the same node [28]. *Location Limited Channels* (LLCs) to bootstrap authentication were first proposed in [17] where nodes are authenticated by 'imprinting' in analogy to ducklings acknowledging the first moving subject they see as their mothers.

In the proposed mechanism, the network is equipped with an infrastructure of authorities and LLCs such that a certificate is delivered over the LLC to a user that has requested it. A device is able to obtain a certificate only upon communicating under a visible monitoring, a *univocal credential*. Such a credential could be either the serial number or MAC address physically assigned to that device. As the physical acquisition of the MAC address is required it cannot be forged. The issued certificate associates the public key of a node (acquired by the authority together with univocal credential) with the hash value of univocal credential and it is signed by the private key of the authority. Thus we propose a mechanism of assignment of unique certificate to each mobile device by a certifying authority that makes it impossible for a node to re-certify itself later with a new univocal credential.

The basic assumption that is made here is that it is not possible to forge some physical attribute of a mobile node (e.g. serial number, MAC address etc.) after it has acquired the same from the certifying authority over the LLC. At any time, the authority thus can ensure that there exists only one valid certificate related to a univocal credential. The univocal credentials are put in encrypted form in the certificates to ensure privacy of the devices.

## 5. THE PROPOSED PROTOCOL

In the proposed scheme, every node in the network monitors the behavior of its neighbors, and if any abnormal action is detected, it invokes an algorithm to determine whether the suspected node is indeed malicious. By 'neighbors' of a node, we mean all the nodes in the network that are one-hop distance from the node. The proposed mechanism builds trust in the network by interactions among some security components running each node. These components are: (i) monitor module, (ii) reputation collector module, (iii) reputation maintainer module, (iv) reputation formatter module, (v) reputation propagator module, and (vi) alarm raiser module. The functionalities of each of these components are described below.

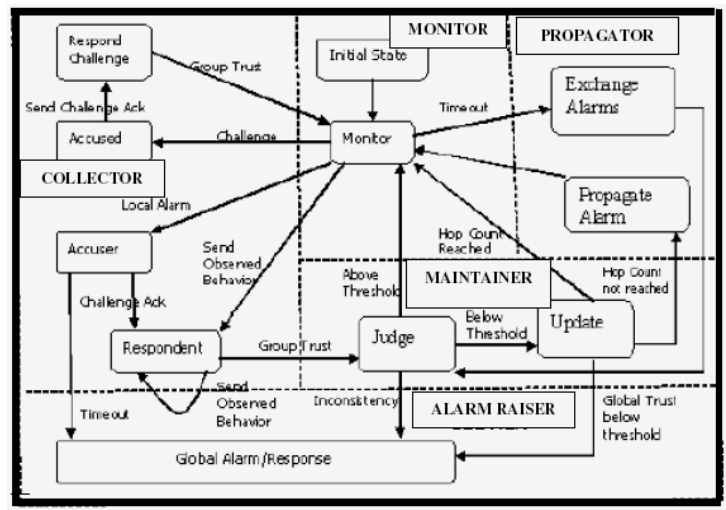

Figure 1. Interactions among the security components in a MANET node

### 5.1. Monitor Module

The monitor module of each node monitors its neighbors by passively listening to their communication. For detecting packet drops and modifications by neighbors, the monitor module of a node randomly copies the incoming packets to its neighbors and checks whether the neighbors really forward the packets with contents unchanged, or drop them, or modify their contents before forwarding them. The collected data is audited by the monitor. The deviation from normal behavior of a neighbor is used as an indicator for the unbiased degree of maliciousness because this is independent of the past behavior of the neighbor node. If this unbiased deviation exceeds a pre-set threshold, the reputation collection module is invoked as discussed below.

### 5.2. Reputation Collector Module

This module invokes a majority consensus algorithm among the neighbors of a node that has been suspected to be malicious. On being activated by its monitor module, the accuser node challenges the suspicious node to verify its behavior as observed by all of its neighbors. The accused node, on receiving the challenge, responds by acknowledging it and sending a *verify_behavior* message to all of its neighbors. The neighbors respond by sending the observed value of degree of maliciousness of the accused. The accused node calculates the group's trust in its behavior using the received values and broadcasts the computed group-trust along with the received responses to all the neighbors. All these messages are encrypted and time-stamped to prevent replay attacks. For computing group trust value from the received responses, any consensus-based scheme may be used. In the proposed scheme, the difference of absolute trust and average degree of maliciousness of the majority of the respondents is used for computing group trust. Majority is taken as the larger of the two groups by partitioning the respondents after comparing their observed degree of maliciousness with a preset threshold.

## 5.3. Reputation Formatter Module

The reputation information of suspected nodes is exchanged between the neighboring nodes in the network (Section 5.2 and Section 5.3) by *rep_mess* messages. A *rep_mess* is an IP datagram with a *reputation header* inserted between the IP header and the data payload. The reputation header consists of three fields: *rep_mess_type*, *node_id*, and *rep_*val. A *rep_mess_type* may be of three types: (i) *rep_request*, (ii) *rep_response*, and (iii) *rep_broadcast*. A *rep_request* message is sent by an accused node when it requests for its reputation value to its neighbors (Section 5.2). A *rep_response* message is used by respondent nodes when they send their observed reputation value about a suspected node. A *rep_broadcast* message is used when a node needs to broadcast the reputation value of a malicious node to all its neighbors.

## 5.4. Reputation Maintainer Module

Every node maintains a global trust state for all maliciously behaving nodes in the network. The trust state is maintained in the form of a *reputation table*. A reputation table consists of two fields: (i) *node_id* and (ii) *rep_val* (Section 5.3). The trust state of a node is updated on when a node receives a new trust certificate. The reputation maintenance module is responsible for verifying the correctness of the group trust certificates received, caching them, and updating the global trust state of the node for which it has received the trust certificate. A node checks the correctness of a certificate by verifying whether response from every neighbor has been correctly considered in computing the group trust and the messages have not been tampered with. This is ensured by application of cryptographic techniques. The contribution of a trust certificate in the final trust value of a suspected node depends on the global trust state of the majority of the neighbors of that node. If the majority in the group observes that the node is acting maliciously, i.e. its trust value is low, the received certificate is propagated to all the neighbors of the accused node (Section 5.5). If the calculated trust value for a node falls below the threshold trust level, a global alarm is raised and the *alarm raiser* module is called on, as discussed in Section 5.6.

For updating the trust value of a node, a cumulative function is used as in (1). In (1), $T_{old}$, $T_{new}$, $T_{certificate}$ stand for the old trust value, new trust value, and the group recommended trust value for a node respectively.

$$(1 - T_{new}) = \alpha.(1 - T_{old}) + \beta.(1 - T_{certificate}) - \delta \qquad (1)$$

$\alpha$ and $\beta$ represent the weightage corresponding to the old trust value and the new trust value of a node respectively. $\delta$ is the trust replenishment factor over time. $\beta$ depends on three factors $\alpha_1$, $\alpha_2$, $\alpha_3$. The parameter $\beta$ can be expressed as in (2).

$$\beta = \alpha_1 . \alpha_2 . \alpha_3 \qquad (2)$$

The parameter $\alpha_l$ is given by (3):

$$\alpha_1 = \frac{\sum_{majority} w_i t_i}{W} \qquad (3)$$

In (3), $w_i$, $t_i$ are the weightage and the trust value respectively of a node belonging to the majority group of the neighbors of the accused node. $W$ is a factor that depends on the size of the network. The factor $\alpha_2$ represents the weightage given to the new trust value computed, and the value of the parameter $\alpha_3$ is obtained using (4).

$$\alpha_3 = \begin{cases} 1 \text{ if } k = 1 \\ \\ 1 \text{ if } k > 1 \end{cases} \qquad (4)$$

## 5.5. Reputation Propagator Module

The reputation propagator module uses mobility of the nodes for propagating trust certificates. Whenever a new trust certificate is issued for a node, it is initially distributed to a subset of nodes that are at the nearest distance (1-hop) from the accused node in the network. This subset of the neighboring nodes is represented as '$F$'. At regular intervals, the neighboring nodes in the network participate in dynamic exchange of certificates. The number of elements in the subset '$F$' determines the effective convergence time of trust information among nodes that are currently and in near future would be the neighbors of the accused node. Intuitively, this can be understood by the fact that initial flooding allows the certificate to be available to the set of nodes that are at 1-hop distance from the accused node and are likely to be first to be the neighbors of the accused. While the accused node moves through the network, every node would receive certificate through flooding and exchange mechanism. The number of hops required to be flooded can be determined dynamically by making neighbors of the accused node send neighborhood information along with observed behavior of the accused node. The certificates are piggybacked on routing packets, and thus involve no communication overhead. This flooding and exchange mechanism enables detection of tampering of packets and provides robustness against packet dropping attacks, as a node can compare a certificate in its local cache with the copy of the same certificate in its neighboring nodes. This scheme is also robust against network partitioning as trust states of the suspected nodes are maintained locally by all the nodes in the network. Moreover, due to group certification scheme, the number of false alarm is also less. As the number of malicious nodes in a network is usually small, the number of trust states to be maintained by the nodes is less. Thus, the scheme requires a very low storage overhead - ideally suited for a resource-constrained MANET node.

## 5.6. Alarm Raiser Module

The alarm raiser module initiates a response action on receiving a global alarm about a malicious node. When a global alarm is raised, the alarm message is flooded across the entire network followed by the invocation of a voting algorithm among the nodes that have recently interacted with the accused node, and final decision about the course of action is taken. Figure 1 depicts the security components in a node and their interactions.

## 6. SECURITY ANALYSIS OF THE PROPOSED SCHEME

In Section 4, it has been discussed that in a MANET, security may be achieved by establishment of PKI and univocal credential of each node. In addition, the redundancy of trust information and the cumulative trust computation function provide robustness in the proposed scheme against message tampering, packet dropping, and false accusation by nodes. Malicious behavior can be shown by the accused node, the accusing and the respondent nodes, or other nodes that do not participate in a particular challenge-response scenario. Security analysis for each of these cases is discussed below.

Case 1: *Accused node is indeed malicious*: In this case the accused node may attempt to perform the following activities:

(i) *Dropping adverse feedback*: A malicious node may try to hide its real behavior by ignoring some adverse feedback in the computation of its group trust. The inclusion of all the received feedbacks from the neighboring nodes in the broadcast message prevents such an attack as any respondent can verify whether its feedback has been correctly considered.

(ii) *Selective broadcast*: A malicious node can use selective broadcast, and drop some adverse responses in its message. However, this is assumed to be not allowed by the network.

(iii) *Tampering of received feedbacks*: Cryptographic mechanisms prevent such attacks. Moreover, a respondent node can detect a tampered message by comparing its local copy of the certificate with that of its neighbors. Since group consensus is majority based, the accused node needs to break in majority of its neighbors to modify the group trust.

Case 2: *Accuser node or the respondent nodes are malicious*. In this case, following situations may arise:

(i) *Deliberate wrong accusation by the accuser node*: An accusing node may make a false accusation against an honest node. However, the attempt may not be successful unless the majority of the neighboring nodes are compromised.

(ii) *False accusation of dropping feedbacks*: If the feedback of an accusing respondent does not change the majority, the malicious node does gain anything by dropping the feedback. However, if the feedback changes the majority, a global alarm is raised. Eventually, either the accusing node or the accused node will be detected as truly malicious. An appropriate intrusion response policy may be implemented to isolate the malicious node.

(iii) *Dropping of trust certificates*: The effect of dropping packets can only slow down the convergence of the trust certificate distribution, because of the use of flooding and exchange mechanism. Mobility of the nodes in the network helps in countering this attack, as more the mobility of the nodes, the faster will be the convergence.

(iv) *Tampering of trust certificates*: Public key cryptography prevents such attacks. However, the algorithm will work even if the cryptographic security is compromised as the redundancy generated by initial flooding can be used to check whether the certificate in the local cache of a node has been tampered with. Only when the majority of the respondent nodes are malicious, these nodes may collude and wrongly accuse a neighbor. The global trust state updation policy proposed in Section 5.4 does not consider more than one certificates received from the same set of nodes as depicted in (4). Also the weightage of a trust certificate depends on the trust state of the recommending nodes (respondents). If the respondents are malicious, the trust certificate issued by them will also have less weightage.

Case 3: *Other nodes are malicious*: If other nodes that do not participate in a challenge-response scenario, i.e., nodes other than the accused node, the accuser node, and the respondent nodes are malicious, then following situations may arise.

(i) *Dropping of trust certificates*: This will only slow down the convergence of the trust certificate distribution. Mobility of the nodes will help in countering this attack. More the mobility of the nodes, the faster will be the convergence.

(ii) *Tampering of trust certificates*: Cryptographic security and redundancy in the network will prevent this activity.

Table 1. Simulation parameters

| Parameters | Values |
|---|---|
| Simulation duration | 1000 seconds |
| Simulation area | 100 m * 100 m |
| Number of mobile nodes | 50 |
| Transmission rage | 200 m |
| Movement model | Random waypoint |
| Maximum speed | 20 m /sec |
| Traffic type | CBR (UDP) |
| No. of network flows | 15 |
| Host pause time | 5 sec |
| Packet flow rate | 2 packets / sec |
| No. of malicious nodes | 5 (10% of total) |

# 7. SIMULATION RESULTS & ANALYSIS

The proposed scheme has been implemented on the network simulator *ns-2* [29] and the performance compared with some existing mechanisms. The 802.11 MAC layer implemented in ns-2 is used for simulation. Nodes with trust value less than 0.4 are taken as *malicious*, those with trust level between 0.4 and 0.9 are assumed be *suspected* and those with trust value greater than 0.9 are assumed to be *trusted*. The trust certificates are exchanged in every one-minute interval. Each node has a buffer capacity of 64 packets with FIFO interface queue. DSR is taken as the routing protocol. The chosen parameters for simulation are presented in Table 1.

The metrics used for evaluation of performance are: (i) false positive rate, (ii) successful detection rate, (iii) total convergence time, (iv) effective convergence time and (v) communication overhead. Total convergence time is the time required for a trust certificate to be propagated to all non-malicious nodes in the network, while the effective convergence time is the time in which all the future and past neighbors of a suspected node will receive its certificate. Effective convergence time is an important metric as the neighboring nodes used in its computation actually participate in determining the degree of malicious ness of a node.

Table 2. Comparative analysis of the performance of different algorithms

| Algorithms / Metrics | LOC | COP | Proposed Algorithm |
|---|---|---|---|
| False positive rate | 90 | 18 | 20 |
| Successful Detection rate | 100 | 85 | 85 |
| Communication overhead (Number of messages) | 120 | 10 | 10 |

The performance of the proposed algorithm is compared with the mechanism presented in [30]. The *local detection algorithm* and the consensus-based *cooperative detection algorithm* of [30] are denoted as LOC and COP respectively in Table 2. It is observed that LOC has a high false positive rate, as in case of congestion it raises a large number of false alarms. However, LOC gives the highest successful detection rate and is used as the baseline of computation on that metric. The drop in performance of COP and the proposed algorithm is because of the fact that only a few malicious nodes will consistently appear in the active path of traffic. While COP will fail to detect these malicious nodes, the proposed algorithm will detect them albeit in a larger time frame. The overheads in terms of false alarms that need to be propagated in the network for a given time interval (160 seconds in simulation) are also compared. The number of group alarms is taken as the measure of the communication overhead. It is observed that for LOC, the overhead of messages is very high as it does not perform any filtering of local alarms before sending them in the network.

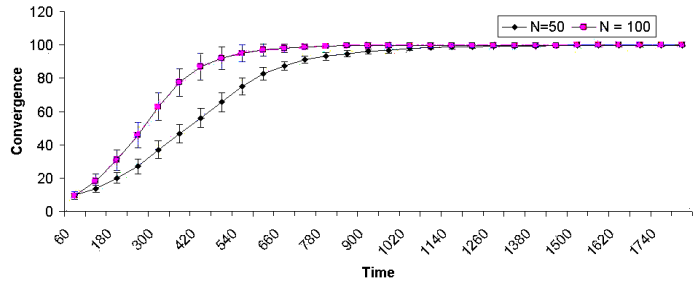

Figure 2. Variation of convergence rate with node density in random waypoint model

Figure 2 shows the variation of convergence rate with different node densities for random waypoint model. Figure 3 shows that the convergence time initially increases with the number of nodes initially flooded for the same node density (1000 in simulation); however, the final convergence time does not differ much.

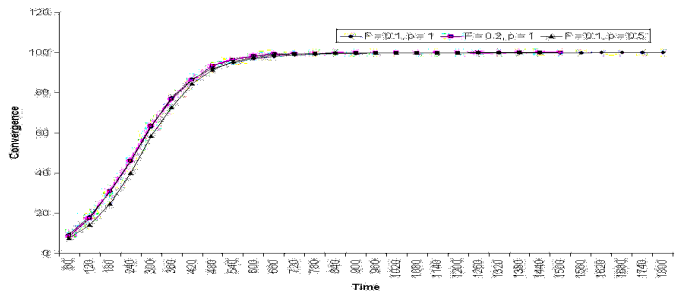

Figure 3. Variation of convergence time with no. of nodes initially flooded in random waypoint

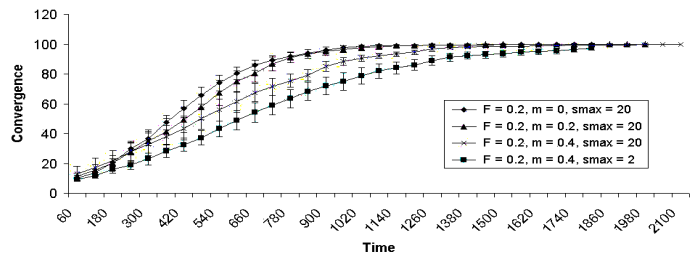

Figure 4. Variation of convergence time with varying no. of malicious nodes in the network

Figure 4 shows the convergence of the algorithm in presence of malicious nodes and different mobility of the nodes. This shows robustness of the proposed algorithm against malicious packet dropping. In Figure 4, the parameters $F$, $m$ and $s_{max}$ represent the percentage of nodes initially flooded, the percentage of malicious nodes in the network, and the maximum speed of the nodes respectively. The results clearly show that the proposed algorithm has no appreciable performance degradation even when 40% of the nodes are malicious. It is also observed that the convergence is faster with higher mobility of the nodes as mobility allows faster interaction among the nodes.

## 8. CONCLUSION

In this paper, an intrusion detection scheme is proposed that enables a routing protocol in MANETs to detect packet dropping attack by a malicious node. In the proposed mechanism, each node independently monitors the packet forwarding behavior of its neighbors. A cooperative mechanism is utilized among the nodes in the same neighborhood for detection of intruders or malicious nodes. The mechanism is simulated in network simulator and the results show that the scheme is highly robust, efficient and has improved performance compared to some existing mechanisms.